\documentclass[twocolumn,amssymb,amsmath,
nofootinbib,tightenlines,showpacs,floatfix,
superscriptaddress,aps,prb]{revtex4-1}
\usepackage{amsfonts}
\usepackage{txfonts} 
\usepackage{amsbsy}
\usepackage{bm}%
\usepackage[colorlinks=true,linkcolor=blue,filecolor=blue,menucolor=yellow,urlcolor=blue,citecolor=blue,anchorcolor=blue]{hyperref}
\usepackage{graphicx}
\usepackage{times} 
\usepackage{dcolumn}

\begin{document}

\title{Transport in Ferromagnet/Superconductor spin valves} 

\author{Evan Moen}
\email{moenx359@umn.edu}
\author{Oriol T. Valls}
\email{otvalls@umn.edu}
\altaffiliation{Also at Minnesota Supercomputer Institute, University of Minnesota,
Minneapolis, Minnesota 55455}
\affiliation{School of Physics and Astronomy, University of Minnesota, 
Minneapolis, Minnesota 55455}

\date{\today}


\begin{abstract}
We consider charge transport properties
in realistic, fabricable, Ferromagnet/Superconductor spin valves 
having a layered structure $F_1/N/F_2/S$, where $F_1$ and $F_2$ denote the ferromagnets,
$S$ the superconductor, and $N$ the normal metal spacer usually
inserted in actual devices. Our calculation is fully
self-consistent, as required
to ensure that conservation laws are satisfied.
We include
the effects of scattering at all the interfaces. We obtain results for the
device conductance $G$, as a function of bias 
voltage, for all values of the angle $\phi$
between the magnetizations of the $F_1$ and $F_2$ layers and a
range of realistic values for the material and geometrical parameters in the sample.  We discuss, in the context of our results for $G$, the relative
influence of all  parameters on the spin valve properties.
We study also the spin current and the corresponding spin transfer torque in $F_1/F_2/S$
structures.
\end{abstract}


\maketitle

\section{Introduction} 

Traditional spin valves\cite{tsyzu} consist of two ferromagnetic materials 
where changing the relative orientation 
of their exchange fields is used to control the transport properties of the heterostructure. They are based on the well-known and much
celebrated\cite{fert} 
Giant Magnetoresistive (GMR) effect. More recently,
it has become possible to
fabricate spin valves by layering
ferromagnetic ($F$) and superconducting ($S$) materials.
In this context, spintronic devices of various kinds\cite{esch,igor,fomi}
have been proposed and considered. The fundamental properties
of such devices arise from the $F/S$ proximity effects\cite{Buzdin2005}.
These effects lead to many new properties. In particular, spin valve
devices, having an  $F_1/F_2/S$ or (more typically
in experimental situations) $F_1/N/F_2/S$, 
where $N$ is a normal spacer,
have
been extensively\cite{kami,wvhg,hvw15,zkhv} studied both theoretically
and experimentally. 
Research on these devices is furthered because, besides their
great scientific interest,
they have possible applications towards
the creation of non-volatile magnetic memory elements. The supercurrents can also be spin-polarized, and this
can then lead to a low energy spin transfer torque that can be used to control the magnetization of nanoscale devices.

Ferromagnetism and $s$-wave superconductivity would
appear to be  incompatible due to the opposite spin structure of their order parameters: the internal fields
in the ferromagnets tend to break the singlet
Cooper pairs. Indeed, although proximity effects 
do exist in $F/S$ heterostructures, they are very different from those 
at $N/S$ interfaces. The exchange field leads
to the Cooper pairs acquiring 
a center of mass momentum\cite{demler} which results
in  damped oscillatory behavior of the singlet pair amplitudes in the $F$ layer regions\cite{Buzdin1990, Halterman2001,Halterman2002}.  
This behavior is 
fundamentally important: it induces oscillations 
in most of the physical properties
of these structures, including the dependence of the 
transition temperature\cite{Buzdin2005}
on the thickness of the various layers. 
It  also drastically changes the 
behavior of transport quantities such as the
the bias dependent conductance, discussed below. 

An even more noteworthy phenomenon arising from 
the $F/S$ proximity effects is that in certain 
$F/S$ heterostructures  triplet correlations may be induced, even though
the $S$ material is an s-wave superconductor\cite{berg86,bvh,Halterman2009}. 
These triplet correlations are necessarily odd in frequency\cite{berezinskii} or,
equivalently, odd in time\cite{hbv,bvh} as required
by the Pauli principle. When the ferromagnetic exchange fields are all aligned
only the $m_z=0$ triplet component can be induced
since $S_z$, the $z$ component of the Cooper pair spin,
commutes with the Hamiltonian. However, when there are two 
or more $F$ layers with  non-collinear exchange fields, as 
can happen for example in $F_1/F_2/S$  structures, 
$S_z$ cannot commute with the Hamiltonian 
and the $m_z=\pm1$ triplet states can also be induced. This is also the
case with a single $F$ layer having a non-uniform magnetization
texture\cite{chiodi,hoprl,ho,gu2015}.
In contrast to the short-range proximity-induced
singlet pair amplitudes, these odd $m_z=\pm1$ triplet states are 
usually long 
ranged\cite{bvermp,Eschrig2008,leksin,Bergeret2007,kalcheim,singh,ha2016} 
in the $F$ layers. Their behavior is also
oscillatory. Because of this, the details
of the geometry of the $F/S$ 
multilayers are crucial to determining their
equilibrium\cite{cko} properties, including
the oscillatory behavior of the transition temperature
with layer thicknesses and with 
the misalignment angle $\phi$ between the 
two $F$ layers in a spin valve\cite{alejandro}.  The
transport properties\cite{wvhg} are also affected. 
As in  a conventional spin valve, the relative exchange field 
orientation of the $F$ layers can have a large effect on the conductance 
of the system. The introduction of triplet correlations can lead to a nonmonotonic dependence of the conductance
on  $\phi$, 
just
as for equilibrium quantities.

Ultimately, all
superconducting proximity effects are governed by Andreev reflection at the interfaces. Andreev reflection\cite{Andreev} is the process of electron-to-hole conversion by the creation or annihilation of a Cooper pair in the superconducting layer. In conventional Andreev reflection, the reflected electron/hole has opposite spin to the incident particle. However, it has been shown\cite{wvhg,linder2009,visani,niu,ji} that in $F/S$ interfaces
triplet proximity effects are correlated with  anomalous Andreev reflection, in which the reflected quasiparticle 
has the same spin as the incident
one. From this, it follows that the transport properties are highly 
dependent on the proper consideration of Andreev
reflection, as has been long recognized 
in both $N/S$\cite{btk,tanaka} and $F/S$\cite{beenakker,zv1,zv2} systems. These effects are particularly important when examining the tunneling conductance in the subgap bias regime where such systems can carry a supercurrent. 
 
In  this paper, we are motivated by the increasing interest in building
actual, practical spin valve structures with potential use as part
of memory elements. We therefore investigate 
the charge 
transport properties of a superconducting spin valve, an
 $F_1/N/F_2/S$ structure 
which includes the normal metal layer spacer, as used in  spin valve devices. 
This normal metal spacer is necessary in experiments 
in order to control 
the relative exchange field of the $F$ layers through the use, for
example, of a pinned and a 
soft ferromagnetic layer, in which the spacer decouples the ferromagnetic layers layers 
(see e.g. Ref.~\onlinecite{alejandro}). 
We will use typical values of the different thicknesses, as  in 
existing and planned devices, and realistic
interfacial scattering between the different layers.
Parameters such as the exchange field and coherence length will
be taken to be in the range relevant to the materials actually used. 
We are particularly motivated to identify the 
relevant experimental 
transport features of actual $F_1/N/F_2/S$ nanoscale systems. Thus, we
 investigate a geometry corresponding
 to experimentally realistic nanopillars with a normal metal layer spacer between two ferromagnetic layers.  These $F/N/F$ layers are grown
 on top of a superconducting substrate. This substrate 
 must be thick enough to allow for the sample to be superconducting:
 its thickness must exceed the superconducting
 correlation length. Furthermore, experimental constraints do not allow for
  perfect interfaces. Although recent developments in fabrication techniques\cite{igor} 
have allowed for very clean interfaces with ballistic transport properties, 
surface imperfections are unavoidable and 
even small interfacial scattering can have a large effect on the transport properties, as we shall see,
 since they affect both ordinary
 and  Andreev scattering. 
We
will use a self consistent solution of the Bogoliubov de Gennes (BdG) equations\cite{degennes} to calculate  the conductance $G$ as a function of bias
voltage for realistic ranges of geometrical and material
parameters, and as a function of the angle $\phi$. Temperature
corrections, which we will show to be non negligible, will also be studied. 
The conductance will be obtained from the self consistent
solutions of the Hamiltonian, via a transfer matrix procedure which
makes use of the Blonder-Tinkham-Klapwijk (BTK) method\cite{btk}. 
In some previous calculations\cite{ji,cheng} of the conductance, a
non self-consistent, step-function pair potential has been
assumed. This neglects the very proximity effects which act on the singlet 
and triplet pair amplitudes, and thus the pair potential. In order to 
properly take these into account, one must use a self-consistent calculation of the pair potential. Even more important, only a self-consistent solution 
 can guarantee that the conservation laws are satisfied\cite{wvhg}, as we review 
in Sec. \ref{methods} below. The feasibility of the methods
we use here was demonstrated in previous work\cite{wvhg} on 
simple $F/F/S$ heterostructures
without $N$ spacers or interfacial scattering, at $T=0$.
That work proved that the
self-consistent 
BTK method embedded into a transfer matrix procedure can be used to 
calculate the tunneling conductance as well as the spin transport quantities. Our work presented here exploits these methods
 with a broader focus on realistic experimental
 parameters and sample compositions. 

 Because of the oscillatory nature of the superconducting singlet 
 (and triplet) amplitudes in the $F$ layers, we will see that,
 as expected, the transport results are highly dependent on  the layer 
 thicknesses, as they are on the exchange field. 
 We report on the $\phi$ dependence of the tunneling conductance as the 
 angular spin valve effect of the system. We do so for a variety of thicknesses for the ferromagnetic and normal layers. Furthermore, we investigate the 
 dependence of $G$ on the interfacial scattering strengths at all the  interfaces. The dependencies that we find are, as a rule, 
 nonmonotonic, and therefore straightforward extrapolations
 are not  possible. 
 Our goal is to provide a better understanding on the full range of experimentally relevant results where the interfacial quality cannot be perfectly controlled. From this, not only can one 
 determine how these parameters affect the spin valve effect, but
 one can also provide the approximate set of parameters that can then maximize this effect: this has both experimental and technological importance.
 We investigate also, in a more  restricted
 set of cases, the spin current and spin-transfer torque (STT).
 

After this Introduction, we briefly review our methods (both for
equilibrium and transport calculations) in Sec~\ref{methods}. The results
are presented, chiefly in graphical form, in Sec.~\ref{results}, and discussed
in the proper context. A summary  Sec.~\ref{conclusions} closes the paper.

\begin{figure}
\includegraphics[width=0.45\textwidth] {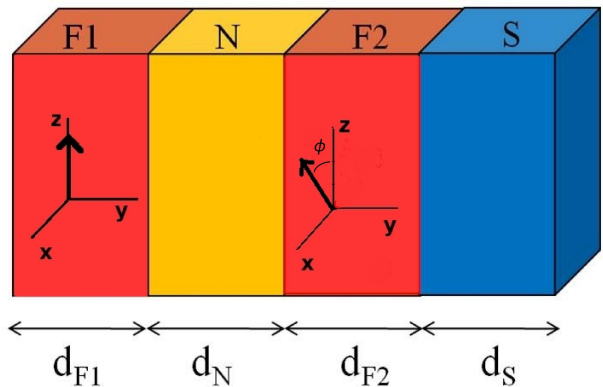} 
\caption{(Color online) Sketch of the structures studied.
The notation for thicknesses of the different layers is indicated,
but the plot is not to scale. The $y$ axis is 
normal to the layers. The magnetizations
of the outer  magnetic layer $F_1$
is along the $z$ axis while  in
$F_2$ it is in the $x-z$ plane, forming
an angle $\phi$ with the $z$ axis, as indicated. 
}
\label{figure1}
\end{figure}

\section{Methods}
\label{methods}
\subsection{The basic equations}

The basic methods and procedures used are straightforward extensions
of those discussed in Ref.~\onlinecite{wvhg} and they need not to be described 
again here.
We merely sketch the main points, in order to establish notation and to make
the paper understandable. The geometry of the system under consideration 
is represented qualitatively in Fig.~\ref{figure1}. The layers are assumed 
to be infinite in the transverse direction. The $y$-axis is normal to
the layers: this somewhat unconventional choice turns out
to be  computationally convenient because only 
the $\sigma_y$ Pauli matrix is complex. The magnetizations of
the outer and inner layers form an angle $\phi$ with each other.

The Hamiltonian appropriate to our system is,
\begin{eqnarray}
\label{ham}
{\cal H}_{eff}&=&\int d^3r \left\{ \sum_{\alpha}
\psi_{\alpha}^{\dagger}\left(\mathbf{r}\right){\cal H}_0
\psi_{\alpha}\left(\mathbf{r}\right)\right.\nonumber \\
&+&\left.\frac{1}{2}\left[\sum_{\alpha,\:\beta}\left(i\sigma_y\right)_{\alpha\beta}
\Delta\left(\mathbf{r}\right)\psi_{\alpha}^{\dagger}
\left(\mathbf{r}\right)\psi_{\beta}^{\dagger}
\left(\mathbf{r}\right)+H.c.\right]\right.\nonumber \\
&-&\left.\sum_{\alpha,\:\beta}\psi_{\alpha}^{\dagger}
\left(\mathbf{r}\right)\left(\mathbf{h}\cdot\bm{\sigma}
\right)_{\alpha\beta}\psi_{\beta}\left(\mathbf{r}\right)\right\},
\end{eqnarray}
where $\Delta(\mathbf{r})$ is the pair potential and 
$\bf{h}$ is the usual Stoner field, which we take
 to be along the $z$ axis (see Fig.~\ref{figure1}) inside the outer magnet $F_1$, while forming an angle $\phi$ with the $z$ axis
 in the $x-z$ plane inside the inner magnet $F_2$. 
We assume 
$h_1=h_2\equiv h$ since in most experiments the same material
is employed. 
The field  vanishes in the superconductor $S$ and in 
the normal spacer $N$.  ${\cal H}_0$ is the single particle
Hamiltonian, which we will take to include the interfacial scattering
as explained below.
Performing a generalized Bogoliubov transformation
in the usual way, with the phase conventions
of Ref.~\onlinecite{wvhg},  and taking advantage of the quasi one dimensional
geometry one can recast the eigenvalue
equation corresponding to the Hamiltonian given
by  Eq.~\ref{ham} as:

\begin{align}
&\begin{pmatrix}
{ H}_0 -h_z&-h_x&0&\Delta \\
-h_x&{ H}_0 +h_z&\Delta&0 \\
0&\Delta&-({H}_0 -h_z)&-h_x \\
\Delta&0&-h_x&-({H}_0+h_z) \\
\end{pmatrix}
\begin{pmatrix}
u_{n\uparrow}\\u_{n\downarrow}\\v_{n\uparrow}\\v_{n\downarrow}
\end{pmatrix} \nonumber \\
&=\epsilon_n
\begin{pmatrix}
u_{n\uparrow}\\u_{n\downarrow}\\v_{n\uparrow}\\v_{n\downarrow}
\end{pmatrix}\label{bogo},
\end{align}
with the $u_{n \sigma}$ and $v_{n\sigma}$ being the usual position and 
spin dependent quasiparticle and quasihole amplitudes involved in 
the transformation. We use units such that $\hbar=k_B=1$.
The quasi one dimensional Hamiltonian
is $H_0=-(1/2m)(d^2/dy^2)+\epsilon_\perp -E_F(y)+U(y)$ where
$\epsilon_\perp$ is the transverse energy, (so that 
the above Eq.~(\ref{bogo}) is a set of decoupled equations, 
one for each $\epsilon_\perp$), 
$E_F(y)$ is the layer dependent width of the band:
$E_F(y)=E_{FS}\equiv{k_{FS}^2}/{2m}$ in the $S$ layer and
$E_F(y)=E_{FM}$ in the $F$ layers.
We define a mismatch parameter\cite{bv} $\Lambda$  as $E_{FM}\equiv\Lambda E_{FS}$. 
$U(y)$ is the interfacial scattering.  We take
this  scattering, due to unavoidable surface roughness
at the interfaces, to be spin-independent and of the form
$U(y)=H_1\delta(y-d_{f1})+H_2\delta(y-d_{f1}-d_N)+H_3\delta(y- d_{f1}-d_N-d_{f2})$. 
The dimensionless parameters $H_{Bi}\equiv H_i/v_F$, where $v_F$ is
the Fermi speed in $S$, conveniently characterize the
strength of the delta functions.

All calculations must be performed self-consistently, otherwise
a large part of the proximity effect is eliminated from the
problem. As previously shown\cite{wvhg,bagwell,sols2,sols}, and as reiterated in Section \ref{conservation},  it is paramount
to perform the transport calculations self-consistently:  not doing so
jeopardizes the law of conservation of change\cite{baym}. The self
consistency condition is:

\begin{equation}
\label{del}
\Delta(y) = \frac{g(y)}{2}{\sum_n}^\prime
\bigl[u_{n\uparrow}(y)v_{n\downarrow}^{\ast}(y)+
u_{n\downarrow}(y)v_{n\uparrow}^{\ast}(y)\bigr]\tanh\left(\frac{\epsilon_n}{2T}\right), \,
\end{equation}
where 
the sum is over all the eigenvalues and
the prime in the sum denotes, as usual, that the sum is limited to states
with eigenenergies within a cutoff $\omega_D$ from the Fermi level.
The superconducting coupling constant $g(y)$, in
the singlet channel, 
is nonvanishing in $S$ only.
Self consistency is achieved by starting with a suitable
choice of $\Delta(y)$ and iterating Eqs.~(\ref{bogo}) and (\ref{del})
until the input and output values of $\Delta(y)$ coincide.
The thermodynamic quantities can then be derived from the wave functions.
The transition  temperature itself can be most conveniently
obtained by linearization of Eq.~(\ref{del}) and an 
efficient eigenvalue technique\cite{zkhv, bvh} as in previous\cite{alejandro}
work.

\subsection{Transport: the BTK method and self-consistency}
\label{BTK}

After the self consistent $\Delta(y)$ function has been obtained as
reviewed above, one can proceed with the calculation
of the transport properties.
There are no fundamental difficulties in extending the 
self consistent\cite{wvhg} BTK 
method\cite{btk} to the case where an extra $N$ layer and interfacial
scattering exists. This is because the only nontrivial
part of the transfer matrix procedure is that which deals with
the self consistent pair potential inside $S$ and this is 
extensively discussed in previous\cite{wvhg} work. For the rest, one has 
of course additional matching equations at the two added interfaces.
The matching equations are of the same basic form as those
found previously\cite{wvhg} except
for the interfacial scattering, which requires, as in elementary
situations, a modification of the derivative continuity condition.
Again, it is not necessary to discuss here these relatively elementary questions, 
although care is required to include them correctly in the
computations. We confine ourselves to the minimum necessary to 
make the notation clear. 

For an incident particle
with spin up the wavefunction in ${F_1}$ is:
\begin{equation}
\label{f1waveup}
\Psi_{F1,\uparrow}\equiv\begin{pmatrix}e^{ik^+_{\uparrow1}y}+b_{\uparrow}e^{-ik^+_{\uparrow1}y}
\\b_{\downarrow}e^{-ik^+_{\downarrow1}y}
\\a_{\uparrow}e^{ik^-_{\uparrow1}y}
\\a_{\downarrow}e^{ik^-_{\downarrow1}y}\end{pmatrix}\enspace\enspace.
\end{equation}
where we have include the appropriate amplitudes for
the ordinary and Andreev reflection processes, which we must calculate.
If the incident particle has spin down, the corresponding wavefunction
in ${F_1}$ is
\begin{equation}
\label{f1wavedown}
\Psi_{F1,\downarrow}\equiv\begin{pmatrix}b_{\uparrow}e^{-ik^+_{\uparrow1}y}
\\e^{ik^+_{\downarrow1}y}+b_{\downarrow}e^{-ik^+_{\downarrow1}y}
\\a_{\uparrow}e^{ik^-_{\uparrow1}y}
\\a_{\downarrow}e^{ik^-_{\downarrow1}y}\end{pmatrix}\enspace\enspace.
\end{equation}
with appropriate amplitude coefficients, numerically
different from those for the spin up
incident particle.  One has, in the above equations:
\begin{equation}
\label{wavevector}
k^{\pm}_{\sigma 1}=\left[\Lambda(1-\eta_{\sigma}{h}_1)\pm{\epsilon}-{k_\perp^2}\right]^{1/2},
\end{equation}
where $\eta_\sigma \equiv 1(-1)$
for up (down) spins, and $k_\perp$ is the length of the wavevector
corresponding to energy $\epsilon_\perp$. All
wavevectors are understood to be in units of $k_{FS}$ and all
energies in terms of $E_{FS}$. 

All of the amplitudes  are then determined from the transfer
matrix procedure discussed in Ref.~\onlinecite{wvhg}, 
where the self-consistent
pair potential determines the wavevectors in the S layer. The transfer matrix matches the continuity conditions for each layer. The outcome of the calculations includes the reflection
amplitudes $a_\sigma$
and $b_\sigma$ of the incoming wavefunctions for the different
(ordinary and Andreev, spin up and spin down) reflection processes. From
these the conductance is extracted as explained below.

\subsection{Conservation laws and conductance}
\label{conservation}

In transport calculations great care has to be taken not
to violate\cite{baym} the conservation laws. Consider
the equation for  charge density $\rho({\mathbf r},t)$ which
arises from the Heisenberg equation:

\begin{equation}
\label{cons}
\frac{\partial}{\partial t}\left\langle\rho({\mathbf r})\right\rangle
=i\left\langle\left[{\cal H}_{eff},\rho({\mathbf r})\right]\right\rangle.
\end{equation}

We are considering here steady state situations, so the time
derivative vanishes and we simply should have a zero a divergence condition
for the current.
In our quasi two dimensional geometry, the only non-vanishing component
of the current is $j_y$, and it depends only on $y$. Hence we need 
to ensure that $\partial j_y/\partial y=0$. Upon computing the commutator 
in the right side of Eq.~(\ref{cons}) under these conditions we find, however: 

\begin{equation}
\frac{\partial j_y(y)}{\partial y}= 2e {\rm Im}\left\{\Delta(y)\sum_n\left[u_{n \uparrow}^* 
v_{n \downarrow}+u_{n\downarrow}^*v_{n\uparrow}\right]\tanh\left(\frac{\epsilon_n}{2T}\right)\right\}  
\label{currentuv}
\end{equation}

In  transport calculations the wavefunctions cannot be taken to be real,
as is possible for the evaluation of static quantities in a current-free 
situation. Hence it is not necessarily true that the right side of 
Eq.~(\ref{currentuv}) will vanish. However, it is easy to  
see\cite{sols2,hvw15} that it will be identically zero when the
self consistency condition Eq.~(\ref{del}) is satisfied. Therefore, the
importance of performing the calculations self consistently, despite the
computational simplifications inherent to non-self-consistent methods, 
cannot be overemphasized.

\subsection{Extraction of the conductance}
\label{extraction}

From the results of the previous subsection, one can extract
the conductance. 
The current is related to the applied bias\cite{btk} 
$V$ via the expression: 
\begin{equation}
\label{totalcurrent}
I(V)=\int G_0(\epsilon)\left[f\left(\epsilon-eV\right) 
-f\left(\epsilon\right)\right]d\epsilon,
\end{equation}
where $f$ is the Fermi function. 
The 
bias dependent tunneling conductance is
$G(V)=\partial I/{\partial V}$.
The function $G_0$ in Eq.~(\ref{totalcurrent}) is the
conductance in the low-$T$ limit or, more generally,
the conductance obtained by replacing the derivative
of the Fermi function by a $\delta$  function. 
It is related to the scattering amplitudes by:
\begin{align}
\label{conductance}
&G_0(\epsilon,\theta_i)=\sum_\sigma P_\sigma G_{\sigma }(\epsilon,\theta_i) 
\\\nonumber
&=\sum_{\sigma}P_{\sigma}\left(1+\frac{k^-_{\uparrow 1}}{k^+_{\sigma 1}}|a_{\uparrow }|^2
+\frac{k^-_{\downarrow 1}}{k^+_{\sigma 1}}|a_{\downarrow }|^2
-\frac{k^+_{\uparrow 1}}{k^+_{\sigma 1}}|b_{\uparrow }|^2
-\frac{k^+_{\downarrow 1}}{k^+_{\sigma 1}}|b_{\downarrow }|^2\right), 
\end{align} 
in the customary  
natural units of conductance $(e^2/h)$. In Eq.~(\ref{conductance}) the 
different $k$ symbols are as defined in Eq.~(\ref{wavevector}). The angle
$\theta_i$ is the angle of incidence: for spin up it is
given by $\tan \theta_i=(k_\perp/k^+_{\uparrow1})$, and similarly
for spin down. Thus
one has $\theta_i=0$ for the forward conductance. The factors
$P_\sigma\equiv (1-h_1\eta_\sigma)/2$ are included to take into
account the different density of  incoming spin up and spin down states.
The energy dependence of $G(\epsilon)$ arises from the applied bias voltage $V$.
It is customary and convenient
to measure this bias in terms of the dimensionless quantity 
$E\equiv eV/\Delta_0$ where $\Delta_0$ is the value of the order parameter in
bulk $S$  material. We will refer to the dimensionless bias dependent
conductance simply as $G(V)$ or $G(E)$  usually omitting the angular argument.

One can not always assume that the experiments
are performed in the low $T$ limit. 
At finite temperature there are two sources of $T$ corrections. The first
and more obvious is that arising from the $T$ dependence of $\Delta(y)$, that 
is, the $T$ dependence of the effective BCS Hamiltonian. This is of
course straightforward to include: one just calculates the
self consistent $\Delta$ at finite $T$ (see Eq.~(\ref{del})
and uses it as input in the transfer matrix
calculations. But there is also a temperature
dependence arising from the Fermi function in Eq.~(\ref{totalcurrent}). If the
temperature  is not too close to $T_{c0}$, the transition temperature of
the bare $S$ material, which sets the overall scale, one can use a 
Sommerfeld type 
expansion. Because the energy scale over which $G(V)$ varies is of order
$\Delta_0$, the relevant expansion parameter is $T/T_{c0}$, not $T/T_F$, and hence
not necessarily negligibly small in all
experimental situations. One finds using elementary\cite{am}
methods:
\begin{equation}
\label{sommerfeld}
G(V,T)=G_0(V)+a_1\left(\frac{T}{\Delta_0}\right)^2 \left(\left.\frac{\partial^2 G(V)}{\partial
\epsilon^2}\right)\right|_{\epsilon=V} + {\cal O}\left(\frac{T}{\Delta_0}\right)^4
\end{equation}
where $a_1$ can be expressed\cite{am} in terms of a Bernoulli number.
Alternatively, one can use the general form:
\begin{equation}
\label{tgen}
G(V,T)=\frac{1}{4 T}\int dV^{\prime}
\frac{1}{\cosh^2[(1/2T)(V-V^{\prime})]}G_0(V^\prime).
\end{equation}
In Eqs.~(\ref{sommerfeld}) and (\ref{tgen}) $G_0(V)$ means the result
of Eq.~(\ref{conductance}) evaluated with the self consistent pair potential
at temperature $T$. The second form turns out to be more useful 
as most relevant  temperatures turn out to be too high for the Sommerfeld expansion.

\subsection{Spin transport}
\label{spintran} 

We will consider also spin transport across the junction. In our
quasi one-dimensional geometry the tensorial spin current
becomes a vector in spin space, while
spatially  it depends only on $y$. Denoting this
vector as $\vec{S}(y)$ it can be written\cite{wvhg} in terms of the
wavefunctions, as:
\begin{equation}
S_i\equiv\frac{i\mu_B}{2m}\sum_\sigma\left\langle \psi_\sigma^{\dagger}\sigma_i\frac{\partial \psi_\sigma}{\partial y}
-\frac{\partial \psi_\sigma^{\dagger}}{\partial
y}\sigma_i\psi_\sigma\right\rangle.
\end{equation}

It is not difficult to write the components $S_i$ in terms of the $u_n$
and $v_n$ wavefunctions. In the $T=0$ limit, the result is:

\begin{subequations}
\label{spincur}
\begin{align}
S_x=&\frac{-\mu_B}{m}{\rm Im}\left[\sum_n\left
(-v_{n\uparrow}\frac{\partial v_{n\downarrow}^{\ast}}{\partial y}
-v_{n\downarrow}\frac{\partial v_{n\uparrow}^{\ast}}{\partial y}\right)\right.\\\nonumber
&\left.+\sum_{\epsilon_\mathbf{k}<eV}\left(u_{\mathbf{k}\uparrow}^{\ast}\frac{\partial u_{\mathbf{k}\downarrow}}{\partial y}
+v_{\mathbf{k}\uparrow}\frac{\partial v_{\mathbf{k}\downarrow}^{\ast}}{\partial y}
+u_{\mathbf{k}\downarrow}^{\ast}\frac{\partial u_{\mathbf{k}\uparrow}}{\partial y}
+v_{\mathbf{k}\downarrow}\frac{\partial v_{\mathbf{k}\uparrow}^{\ast}}{\partial y}\right)\right]\\
S_y=&\frac{\mu_B}{m}{\rm Re}\left[\sum_n\left(-v_{n\uparrow}
\frac{\partial v_{n\downarrow}^{\ast}}{\partial y}
+v_{n\downarrow}\frac{\partial v_{n\uparrow}^{\ast}}{\partial y}\right)\right.\\\nonumber
&\left.+\sum_{\epsilon_\mathbf{k}<eV}\left(u_{\mathbf{k}\uparrow}^{\ast}\frac{\partial u_{\mathbf{k}\downarrow}}{\partial y}
+v_{\mathbf{k}\uparrow}\frac{\partial v_{\mathbf{k}\downarrow}^{\ast}}{\partial y}
-u_{\mathbf{k}\downarrow}^{\ast}\frac{\partial u_{\mathbf{k}\uparrow}}{\partial y}
-v_{\mathbf{k}\downarrow}\frac{\partial v_{\mathbf{k}\uparrow}^{\ast}}{\partial y}\right)\right]\\
S_z=&\frac{-\mu_B}{m}{\rm Im}\left[\sum_n\left(v_{n\uparrow}\frac{\partial v_{n\uparrow}^{\ast}}{\partial y}
-v_{n\downarrow}\frac{\partial v_{n\downarrow}^{\ast}}{\partial y}\right)\right.\\\nonumber
&\left.+\sum_{\epsilon_\mathbf{k}<eV}\left(u_{\mathbf{k}\uparrow}^{\ast}\frac{\partial u_{\mathbf{k}\uparrow}}{\partial y}
-v_{\mathbf{k}\uparrow}\frac{\partial v_{\mathbf{k}\uparrow}^{\ast}}{\partial y}
-u_{\mathbf{k}\downarrow}^{\ast}\frac{\partial u_{\mathbf{k}\downarrow}}{\partial y}
+v_{\mathbf{k}\downarrow}\frac{\partial v_{\mathbf{k}\downarrow}^{\ast}}{\partial y}\right)\right], 
\end{align}
\end{subequations}
where  the first terms in the right side are the spin current components
in the absence of bias. A  static spin transfer current may exist
near the boundary of two magnets with misaligned fields. The above
results are valid at low $T$, 
 we will not consider temperature corrections 
for this quantity.  In the steady state the conservation laws require:

\begin{equation}
\label{spinconserve}
\frac{\partial}{\partial y} S_i= \tau_i,
\enspace\enspace i=x,y,z 
\end{equation}
where $\bm{\tau}$ is the torque $\bm{\tau}\equiv 2\mathbf{m}\times\mathbf{h}$
with $\mathbf{m}$ being the local magnetization $\mathbf{m}=-\mu_B\sum_{\sigma}\langle\psi_\sigma^{\dagger}\bm{\sigma}\psi_\sigma\rangle$. 
The expression for $\mathbf{m}$ in terms of the
wavefunctions is given in Ref.~\onlinecite{wvhg}.

\section{Results}
\label{results}

In this section we present our results. As discussed in the
Introduction, our emphasis is in exploring a range of
values of experimental interest for the relevant parameters. 
This, in addition to
helping us meet our goal of helping experimentalists understand their
data, will keep the discussion within reasonable bounds: otherwise,
with a more than ten-dimensional parameter space to be investigated,
this work would completely lose its focus. 
We do have an extensive and growing database of results for many other
cases.
As mentioned above, we use dimensionless parameters in our plots:
all lengths are given in units of $k_{FS}$ and all energies in units
of $E_{FS}$ except, as already stated, for the bias. Dimensionless lengths 
will be denoted
by capital letters with the appropriate subscript. The units for the
dimensionless barrier height parameters $H_{Bi}$ have been explained
before. Values close to unity would represent a strong tunneling
limit: these would be experimentally very
undesirable as the proximity effects would 
be very small. Zero values represent an ideal interface,
which is unlikely to be attainable experimentally. Since
the first and second interfaces are both between $F$ and $N$ materials,
one can fairly safely assume that these two barrier strengths are similar, 
and we will usually take them to be identical, $H_{B1}=H_{B2}\equiv H_B$. 
In our dimensionless units a field parameter
value of $h=1$ would correspond to a half metal. 
The
results for $G$ presented are for $h=0.145$  a value previously
found adequate\cite{alejandro} in fitting Co static properties in similar 
devices.
As in Ref.~\onlinecite{alejandro} we set $\Lambda=1$, which subsumes some of the
wavevector mismatch effects with the phenomenologial $H_{Bi}$ parameters.  
We will also assume a value of $\Xi_0=115$ for the 
dimensionless correlation length in $S$, a value used in the
same context\cite{alejandro} for Nb.
We will vary the thicknesses of all layers, keeping $D_{F2}$ relatively
small, which is necessary to obtain good proximity effect, and
allowing $D_N$ and $D_{F1}$ to be somewhat larger. As to $D_S$,
the thickness of the superconducting layer, it must
of course be kept above $\Xi_0$: otherwise the
sample tends to become non-superconducting, for rather obvious reasons. 
We will focus here on forward conductance results, which can be obtained
from point probes and involve trends much easier to understand.

\begin{figure}
\includegraphics[width=0.40\textwidth,angle=-90]{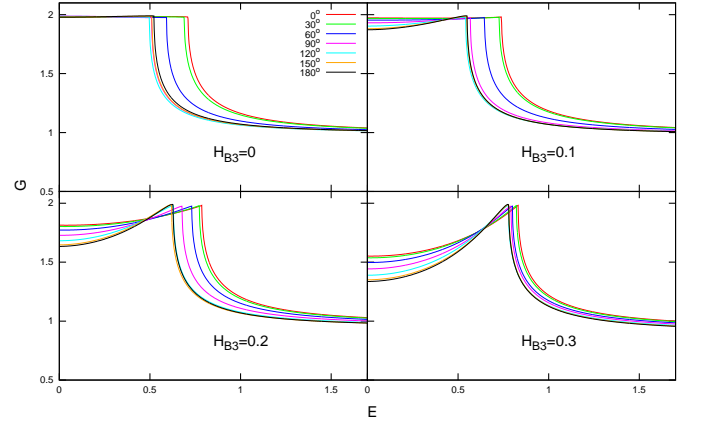}
\vspace{-2mm}
\caption{Effect on the conductance of the barrier between the superconductor
and the inner ferromagnet $H_{B3}$. The four panels show results
for $G$ in natural units, as a function of bias voltage $E \equiv eV/\Delta_0$ 
at seven values of the misalignment angle $\phi$ as indicated in the
legend. The panels correspond to different values of $H_{B3}$ ranging
from 0.0 to 0.3 with
$H_{B1}=H_{B2}\equiv H_B=0$. The thicknesses
are $D_{F1}=20$, $D_{N}=40$, $D_{F2}=12$ and $D_S=180$. The internal
field parameter is $h=0.145$}
\label{figure2}
\end{figure}
\subsection{Barrier effects}

The effects of interfacial scattering are very strong and important.
Recall that even in standard normal-superconductor interfaces the
zerto bias 
conductance (ZBC) can vary between a value of two for a perfect
interface, and an exponentially small value for the tunneling limit. 
One should recall here that even in the case where a certain
barrier parameter vanishes, there is still scattering at the
correspondent interface: this is because it is impossible for the
two Fermi wavectors in the ferromagnets to match the Fermi wavevector 
of either the $N$ or the $S$ materials. This has to be kept in mind
in the discussion below.
 
\begin{figure} 
\hspace*{-4mm}\includegraphics[width=0.40\textwidth,angle=-90]{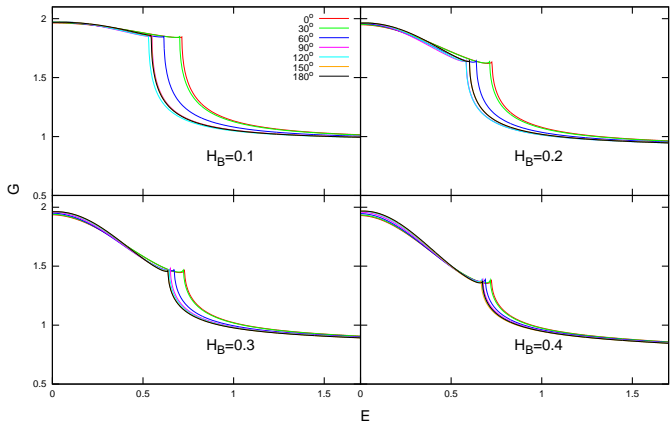}
\vspace{-2mm}
\caption{Effect on the conductance of the barriers between the normal
spacer and the
ferromagnets $H_{B1}=H_{B2}=H_B$. The four panels show results for
the same arrangement as in Fig.~\ref{figure2} and the same geometrical
and field parameters except
in this case 
$H_{B3}$ is held constant and the  value of the barrier parameter at
the other two interfaces is varied between 0.1 and 0.4. 
}
\label{figure3}
\end{figure}
In
Fig.~\ref{figure2} we show the effect of increasing $H_{B3}$ assuming
that the other interfaces have zero interfacial potential, although 
scattering due to wavevector mismatches is present. 
Four values of $H_{B3}$ are studied, one
in  each panel, and curves for seven values of 
the misalignment angle $\phi$ are plotted.
The geometrical parameters 
are $D_{F1}=20$, $D_{N}=40$, $D_{F2}=12$ and $D_S=180$.
The overall trend on increasing $H_{B3}$ is a marked decrease of the
low bias conductance and a much smaller decrease of the high bias limiting
value. The critical bias (CB) is the value of the bias
at which $G$ sharply changes behavior and begins trending towards its
normal state limit. In general, the
critical bias is smaller than unity, and smaller values
are associated with stronger proximity effects since the CB is
associated with the saturated value of $\Delta(Y)$ well inside $S$.
We see that the CB tends to increase with
$H_{B3}$, while the value of $G$ at critical bias (the critical
bias conductance, CBC) remains nearly the same. On the other
hand, the CB is in all cases a strong function of $\phi$,
decreasing as $\phi$ increases, up to about $\phi=100^\circ$ and then flattening, for this geometry. 
The dependence
is less marked at higher barrier values. The 
ZBC however, is monotonically
decreasing  in $\phi$. 
This dependence on $\phi$ is different from that of the CB or CBC, 
and it leads to a 
crossover in the conductance values. Remarkably, this crossover tends to occur with a "nodal" behavior at a single bias value in the subgap region: this can best be seen in the third and fourth panels.
Monotonic behavior in the ZBC also occurs for other values of $D_{F2}$ that we have
studied, but the direction (increasing or decreasing
in $\phi$) is reversed in an oscillatory way: for example the ZBC 
increases with $\phi$ 
at values of $D_{F2}$ of 7 and 10 and again at 16,17. This is one more
example of the multiple oscillatory behavior found  in this problem
and an illustration of how much care one has to take before extrapolating
results.

\begin{figure*}
\includegraphics[width=0.6\textwidth,angle=-90]{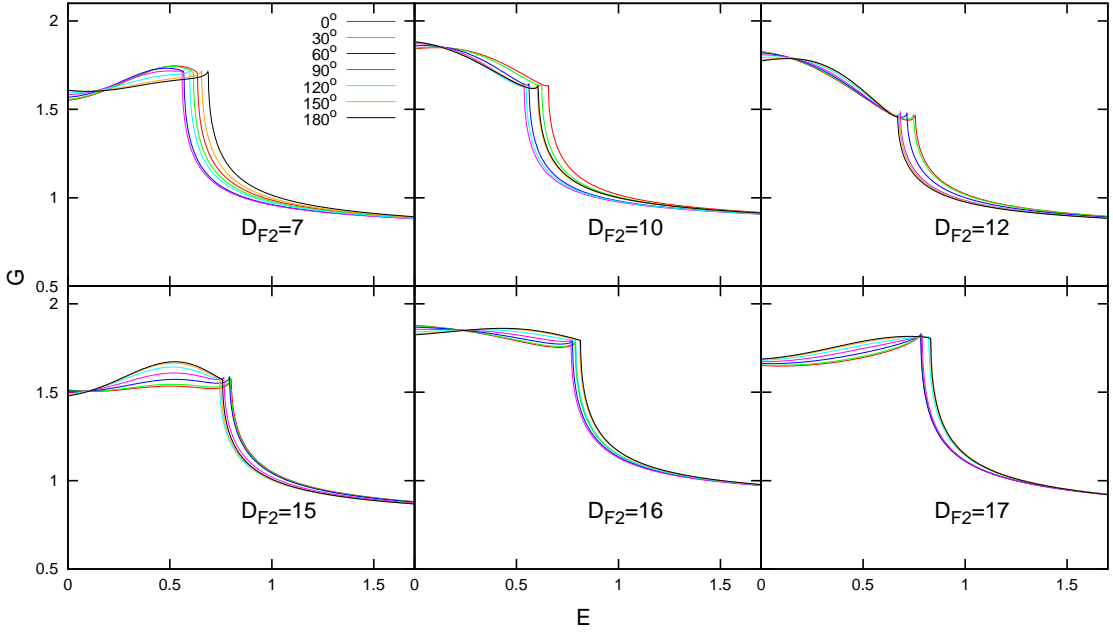}
\vspace{-10mm}
\caption{Effect on the conductance by varying the thickness
$D_{F2}$ of the inner ferromagnetic layer. The values of
the other thicknesses, field, and correlation
length are as in the previous two figures, and the barrier values
are set to 0.3, 0.3, and 0.1 respectively, which are representative of possible experimental values. The six panels show
$G$ vs bias voltage for several angles, at six values of 
$D_{F2}=7,10,12,15,16, $ and $17$. The spin valve effect varies significantly in both the CB and the ZBC.} 
\label{figure4}
\end{figure*}

Next we consider, in Fig.~\ref{figure3}, the effect of increasing
$H_{B1}=H_{B2}\equiv H_B$ while keeping $H_{B3}=0$ at
the $F_2/S$ interface. Again, four 
barrier values are considered, in an arrangement very similar
to that in the previous figure. The effects of interfacial scattering
are now more pronounced. This is not necessarily
due to  the presence of two barriers:
as in well known situations in elementary one-dimensional
quantum mechanics, we find that having more barriers 
does not necessarily lead to less 
transparency. This analogy is imperfect: our system
is not one-dimensional,  there are multiple scattering
mechanisms (interfacial imperfections, wavevector mismatch, Andreev
reflection, etc).
Still, we find that having  two barriers does not always reduce transmission.
A clear example of this can be seen in the ZBC value which, for the chosen
values of $D_{F2}=12$ and $D_N$, is nearly independent of $H_B$. This is because
of resonance-like  behavior  in this geometry. Furthermore, changing the
values of $D_{F2}=12$ and $D_N$ leads to ZBC
behavior more similar to that in Fig.~\ref{figure2}, which we discuss in the next subsection 
in connection with Fig.~\ref{figure6}. 
The behavior of the CB with angle is nonmonotonic, in a way similar to
that found in Fig.~\ref{figure2}. The minimum is now somewhat less shallow,
particularly
at higher $H_B$. At  low bias, $G$ decreases as the bias is increased,
although an upturn does occur as the CB is approached albeit at a lower value of the CBC for increasing $H_B$. This is in contrast to Fig.~\ref{figure2} where the CBC was unaffected by $H_{B3}$.

\subsection{Geometrical Effects}

We have mentioned 
in the previous discussion that the thickness
of the different layers may have a strong and often nonmonotonic effect
on $G$. The thickness of the inner magnetic layer, $D_{F2}$ turns
out to be the
more important of these geometrical variables. 
In the six panels in Fig.~\ref{figure4} we consider increasing
values of $D_{F2}$ while keeping the other geometrical and 
material parameters 
fixed to their values in the previous figures. The three 
interfacial barrier parameters are set to intermediate values (see
the caption).

Consider in detail the first panel, where $D_{F2}=7$. One notices
immediately the reduction in ZBC, as opposed to the results 
for $D_{F2}=12$ in the third panel or to those in the previous figures. The behavior of this
reduction occurs, as has
been mentioned above, in an  oscillatory
manner with $D_{F2}$: it can be seen again at 
$D_{F2}=15$ (fourth panel). In this panel, as in the second and the
fifth, the minimum value of the CB with angle is at $\phi=90^\circ$, and
this minimum is very well marked -- this is an
optimum situation for valve effects. 
The ZBC value depends somewhat on $\phi$
but not in the same way as the CB: hence, the crossing conductance curves 
near a bias
of 0.2.  The second panel exhibits similar behavior, but the ZBC
is markedly higher. On
further increasing $D_{F2}$ to 12 (third  panel) the  CB becomes
monotonic in $\phi$ while the low bias conductance does not change:
indeed the node where the lines cross barely moves. The case $D_{F2}=15$
(fourth panel) is yet different: the CB is larger and there
is a marked ``bump'' in the low bias conductance, the height of which
increases with $\phi$. Resonance in the ZBC is observed again
in the fifth panel, and the angular dependence of the CB returns
to having a marked minimum at $\phi=90^\circ$ although with a weaker
dependence. Furthermore, the node noticeably moves to a higher bias value. Finally, at $D_{F2}=17$ (last panel) the ZBC
drops again, the angular dependence of the CB is reversed, and the node disappears.
Thus we see that the thickness of the inner magnetic layer is
a very important variable in determining the conductance properties.

\begin{figure}
\includegraphics[width=0.35\textwidth,angle=-90]{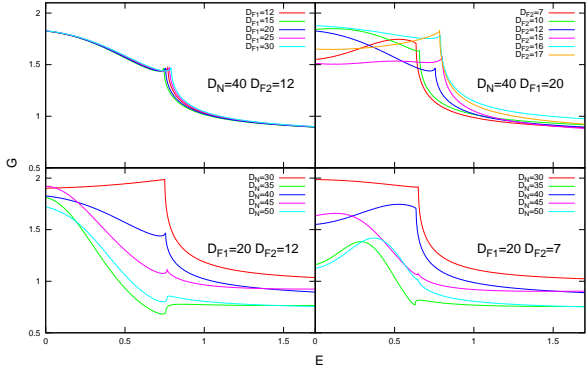}
\vspace{-2mm}
\caption{Effects of varying $D_{F1}$ or $D_N$, compared with
dependence on $D_{F2}$. All panels
are for $\phi=0$,  barrier values
of  0.3, 0.3, and 0.1 and  the field parameter, correlation
length, and $D_S$  are as in Figure \ref{figure2}.
The first two panels contrast
the effect on the conductance of varying the thickness
$D_{F1}$ of the outer ferromagnetic layer with $D_{F2}$ of the inner ferromagnetic layer. In the first panel,  $D_{F1}$
is varied, as indicated in the legend, at $D_{F2}=12$,
while in the second one $D_{F2}$  is varied at $D_{F1}=20$.
The last two panels show the effect of varying
$D_N$ at $D_{F1}=12$ and $D_{F2}=7$ respectively. 
The dependence of the results on $D_{F1}$ is much weaker than
that on $D_{F2}$ or $D_N$. Both $D_{F2}$ and $D_N$ have a large impact on the ZBC, meanwhile $D_{F2}$ has a much larger effect on the CB. }
\label{figure5}
\end{figure}

On the other hand, the effect of varying $D_{F1}$, the thickness
of the outer ferromagnetic layer,  is much weaker
than that of varying $D_{F2}$. This
is illustrated in the first two panels
of Fig.~\ref{figure5}. There we display, in each panel,
results for $G$ at fixed $\phi=0$.  In the first panel
we do this for several values of
$D_{F1}$ ranging from 12 to 30 and, in the second panel, for $D_{F2}$  
values  from 7  to  17 at fixed $D_{F1}$. In both panels
$D_N=40$. Barrier 
heights and other parameters
are as in  Fig.~\ref{figure4}.  
The difference is
obvious: while in the first panel the results barely 
change (although the change is nonmonotonic),  in the second 
 one every relevant quantity 
(CB, ZBC, high
bias and low bias behaviors etc) changes, in obvious
and very strongly  nonmonotonic ways.
Thus, in the fabrication process, the precise thickness of $D_{F1}$ is
less critical than that  of $D_{F2}$. 
As to the normal spacer thickness, in the last two panels of  
Fig.~\ref{figure5}
we 
consider the dependence 
of $G$ on $D_N$. 
We again plot  $G$ at fixed $\phi=0$ for several values of $D_N$ at two values
of $D_{F2}$ (see caption). One can see that while quantities such as the
CB do not depend very much  on $D_N$, the low and high bias behaviors 
vary quite appreciably overall, the former rather dramatically.
Hence we conclude that
$D_{F2}$ is the crucial geometrical parameter in the problem, followed
in importance by $D_N$ and with $D_{F1}$
being much less relevant.

\begin{figure}
\includegraphics[width=0.35\textwidth,angle=-90]{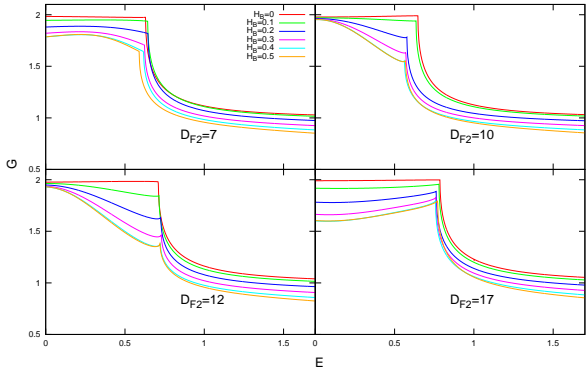}
\vspace{-2mm}
\caption{Combined effect of $D_{F2}$ and barriers. The behavior
at fixed $\phi=0$ and $H_{B3}=0$ is studied. Each of the four
panels
corresponds to a fixed value of $D_{F2}$: 7, 10, 12, and 17
and the curves correspond to
values of $H_{B1}=H_{B2} \equiv H_B$ as indicated in the legend. A
nonmonotonic feature in the ZBC is observed as a function of $D_{F2}$, 
owing to the oscillatory behavior of the Cooper pairs.
}
\label{figure6}
\end{figure}

Careful examination of the above results yields insights on the combined
effects of interfacial scattering and on geometry, particularly on 
$D_{F2}$: how geometry and interfacial strength are 
related follows ultimately from 
the oscillatory nature of the Cooper pairs and
from quantum mechanical interference. We now display, in Fig.~\ref{figure6}, 
these
combined effects in a more direct way. As in Fig.~\ref{figure5}
we study results for fixed $\phi=0$. 
We consider four values of $D_{F2}$, one in
each panel, ranging from 7 to 17, and plot results for several values
of $H_B$ at $H_{B3}=0$. In the first panel we see a large
and monotonic 
dependence on $H_B$ of the entire conductance dependence. In the next
case shown, $D_{F2}=12$, the ZBC depends only very weakly on $H_B$. In the
next panel, the spread in the ZBC with $\phi$ increases 
somewhat, as compared to the previous panel, and it does so 
even more in the last panel. 
This resonance-like behavior is not the same as in the one-dimensional two 
barrier problems in basic quantum mechanics, where a resonance feature is observed in the transmission coefficients as  a function of the 
distance between the barriers. This analogy might 
apply better to $D_N$, but not to the inner ferromagnetic thickness $D_{F2}$. Instead, this resonance is due to the oscillatory behavior of the Cooper pairs. We see then that certain values of $D_{F2}$
make the system, or at least its ZBC, partly ``immune'' to the effects
of fairly high surface barriers. Although this holds only to a limited
extent, it may be worthwhile to attempt to exploit this effect to
palliate the existence of unfavorable interfaces with unavoidably large
scattering.

\begin{figure}
\hspace*{-25mm}\includegraphics[width=0.55\textwidth,angle=-90]{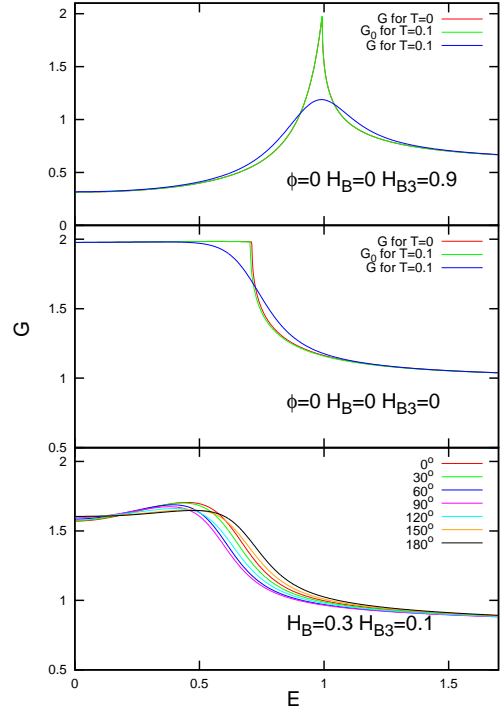}
\vspace{-2mm}
\caption{Temperature dependence of the conductance.  In the 
first two panels we  consider $G$
at fixed $\phi$. The 
thicknesses and fields are as in Fig.~\ref{figure2}. 
Temperatures  $T=0.1$, in units of $T_{c0}$, are compared
to $T=0$ results. 
The result of including only $G_0$, the correction to $G$ arising 
from the $T$ dependence of $\Delta(y)$ is also shown. The first panel 
is for  a very high barrier ($H_{B3}=0.9$) between $S$ and $F_2$
and $H_{B1}=H_{B2}=0$, while in the second all 
$H_{Bi}=0$. The $G_0$ result at $T=0.1$ is nearly identical to the
$G$ at $T=0$.
The last panel illustrates 
(for the same values as the first panel in Fig.~\ref{figure4}), 
a case where
the CB varies very nonmonotonically 
with angle, and shows how little this behavior is affected by $T$.  }
\label{figure7}
\end{figure}

\begin{figure*}
\includegraphics[width=0.7\textwidth,angle=-90]{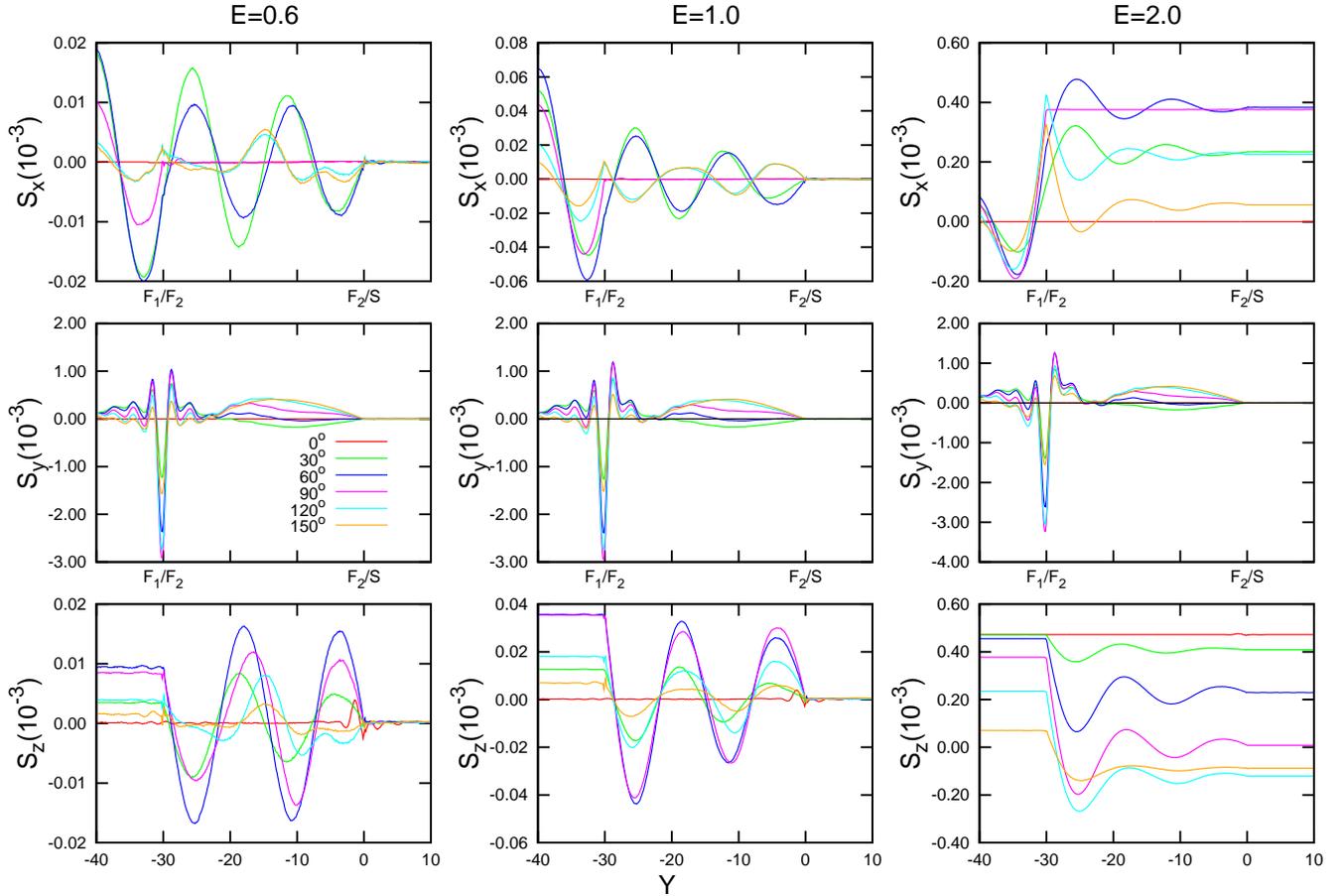}
\caption{The three components of the spin
current are shown as a function of $Y$ for several values of $\phi$,
as indicated, and three values of the bias voltage. We
have $h=0.1$, $D_{F1}=D_{S}=250=5 \Xi_0$, $D_{F2}=30$, $D_N=0$.
Only the central region of $Y$ is plotted: $Y=0$ is at the $F_2/S$
interface. All components of the spin current are zero for 
$\phi=180^\circ$. 
} 
\label{figure8}
\end{figure*}

\begin{figure*}
\includegraphics[width=0.7\textwidth,angle=-90]{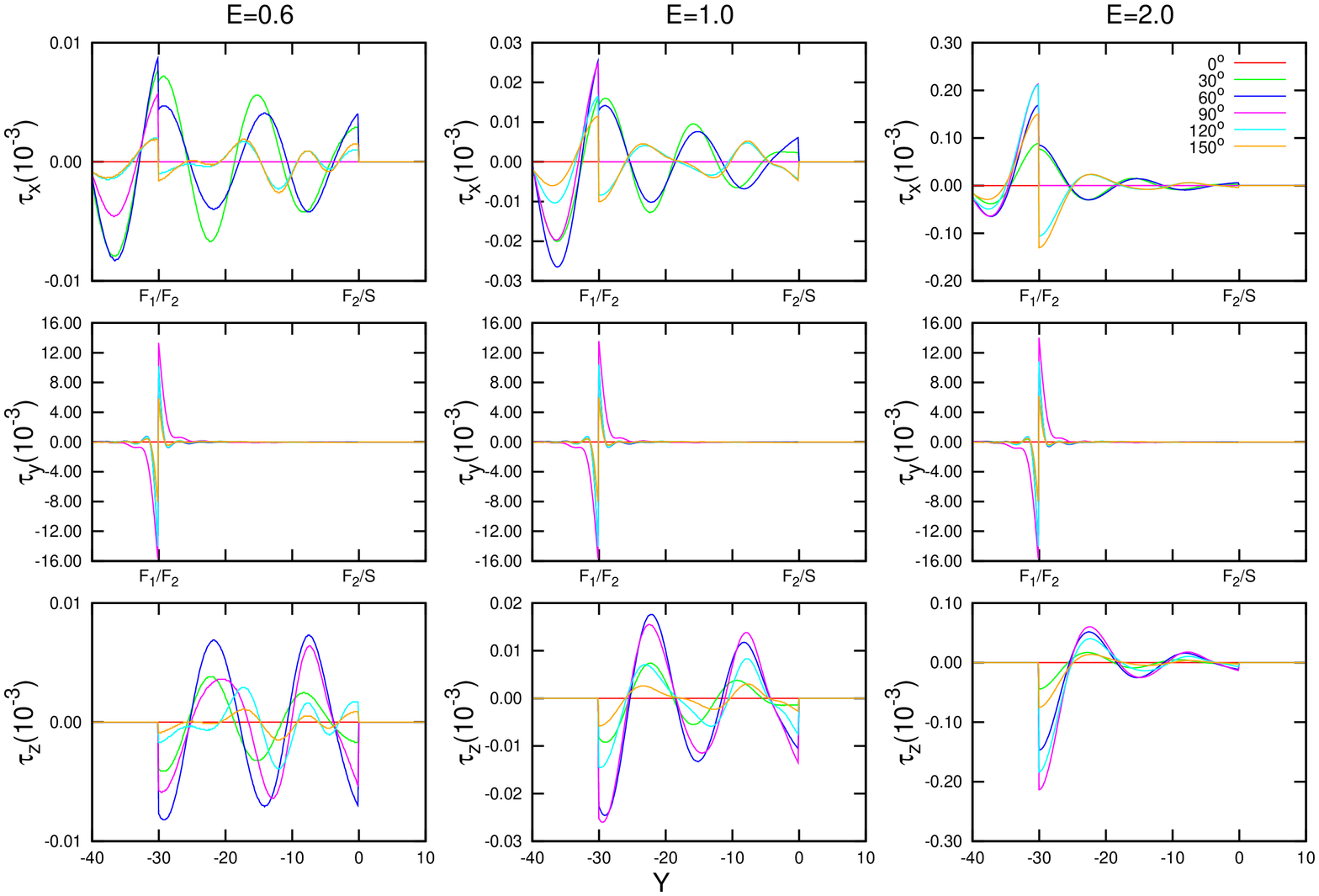}
\caption{The three components of the spin transfer torque plotted for the same situation
as in the previous figure. The torque is 
identically zero for $\phi=0$ and $\phi=180^\circ$. The discontinuities
at the interface reflect those of 
 the internal fields.} 

\label{figure9}
\end{figure*}
 
\subsection{Temperature dependence}
 
Experiments in these systems are not performed at zero temperature,
nor, in practice, at ultralow $T$. Therefore the
influence of $T$ must be examined. There are two transition temperatures
to consider: the transition temperature $T_{c0}$ of pure bulk $S$
material, and the transition temperature $T_c$ of the device, which is
typically considerably lower. In our discussion we will use a
dimensionless temperature $T$ in units of $T_{c0}$ since $T_c$ varies
as the geometry is changed.

As explained in Sec.~\ref{extraction} one has to consider 
two sources of $T$ dependence. The first
is 
that arising from the self-consistent pair potential, $\Delta(y)$, 
that is, the $T$ dependence in the effective Hamiltonian. 
This leads to the function $G_0$ defined below Eq.~(\ref{totalcurrent}) 
and in Eq.~(\ref{conductance}) being $T$ dependent. 
The second is
that originating in  the Fermi functions
in Eq.~(\ref{totalcurrent}). As  discussed in connection with
Eq.~(\ref{sommerfeld}) the latter is not negligible since the scale
of the variation of $G$ with bias is $\Delta_0$, not the Fermi energy.
We have found that, in practice,  Eq.~(\ref{tgen}), which
is not dependent on any expansion, is much more useful
than the Sommerfeld method in the relevant temperature range. This
is because the conductance has large, and even discontinuous
derivatives, which the Sommerfeld expansion does not handle well.

Representative results are shown in Fig.~\ref{figure7}. In the first two
panels we consider a fixed $\phi=0$ and we show results for $G$ both
at $T=0$ and at a reduced temperature $T=0.1$. Since for the size ranges 
considered 
in this section we have found that
$T_c/T_{c0}$ values are in the 0.5 to 0.6 region, these correspond
to $T/T_c$ of about 0.2. The first panel shows results in a strong tunneling
limit regime, with high barriers, and the second for  zero barrier 
heights. Plots of $G_0$ , i.e. the 
results obtained by using the $\Delta(y)$ correction only are also included:
these are obviously inadequate in both cases, and the full result is needed. 
We have found this to be invariably the case except at
unrealistically low $T$. The
overall effect of the temperature is, otherwise, that of 
rounding up and softening
the sharp features of the low $T$  results. A consequence of this is
that at finite $T$ one has to redefine more carefully the CB as the
bias value at which $G$ has a peak or a high derivative. The proper
redefinition is the bias value at which $G$ varies fastest.

In the third panel of 
Fig.~\ref{figure7}, we replot $G$ for the same case considered in the first 
panel of Fig.~\ref{figure4}, 
which, as  we have remarked before, shows good spin valve effects
in its CB properties, 
but now at $T=0.1$ instead of at zero temperature.
The two results should be carefully compared.
We see that while the curves are now much smoother the behavior of
the different features with angle are robust. In particular the 
sharp minimum
of the critical bias at $\phi=90^\circ$ remains unchanged. We have found
this to be the the situation in all the cases we have checked.  Hence,
spin valve properties are only weakly dependent on $T$.

\subsection{Spin Currents}
\label{spincurrents} 

We present here some results for the spin current and 
the spin transfer torque. 
We restrict ourselves to the case where there is no spacer, and the
barrier parameters are zero. 
However, we consider in this paper a range of bias voltages
and all values of the 
angle $\phi$. 
Very limited 
results for only $\phi=90^\circ$ value were given in Ref.~\onlinecite{wvhg}.
We use units such that $\mu_B=1$ and take $h=0.1$. We consider
a superconductor thickness of five times the coherence
length ($D_S=250=5\Xi_0$) so that the saturated value of $\Delta(y)$ is essentially the same as the bulk $S$
value $\Delta_0$. 
We assume a rather thick $F_1$ layer ($D_{F1}=250$) 
while $D_{F_2}=30$. 

The main quantities  we will focus on  are the three
components of the spin currents and of the spin transfer 
torques (STT) as a function of position. For the charge current,
the conservation law entails that the current is independent of
position. 
But for spin, the derivative of the current is the STT 
(see Eq.~(\ref{spinconserve})) and the latter quantity
is of great physical interest. As usual\cite{hbv,wvhg}  we normalize
$\mathbf{m}$  to $-\mu_B(N_\uparrow+ N_\downarrow)$. The normalization
for the spin current follows from these conventions. There are two
alternative methods to calculate the spin currents: one is directly
from the expressions in Eqs.~(\ref{spincur}). The other method is
to calculate the torque first, from the expression below Eq.~(\ref{spinconserve})
and then  integrate over the $y$ variable. The two methods 
agree when the calculations are done self consistently,
as was conclusivelly shown in Ref.~\onlinecite{wvhg}. 
The second method
is computationally much easier, but it yields results only up to
a constant of integration. We have therefore used the direct method: it
requires obtaining wavefunction results over a very fine mesh, so that 
the derivatives in Eq.~(\ref{spincur}) can be calculated to sufficient
accuracy.

In the following discussion it is well to recall the meaning of 
the indices and coordinates. The spin current is in general a tensor, each element
having two indices, one corresponding to the spatial components and the other
to spin. In a quasi-dimensional geometry,  the only spatial component
is in the $y$ direction, normal to the layers 
in our convention (see Fig.~\ref{figure1}). The spin current is then
simply a vector in spin space: the indices in $S_i$ denote spin components,
with all transport being in the spatial $y$ direction. Recalling
Eq.~(\ref{spinconserve}) and the definition of the torque 
$\mathbf{\tau}=2\mathbf{m}\times\mathbf{h}$ 
we see that $\tau_y$ tends to twist the magnetization
in the  plane of the layers, but of course it can only do so in
regions near the interfaces, where $\mathbf{m}$ and $\mathbf{h}$ 
are not parallel due to magnetic proximity effects. We also see that each component of the torque vanishes in the $S$ layer where the internal field parameter $h$ is zero. 

We can now discuss the plots in Figs.~\ref{figure8} and \ref{figure9}. These
two figures show results 
for the three components of the spin current and of the STT respectively, each under the same conditions (see captions). These quantities are shown for three
values of the bias, $E$, ranging from  below to well above $\Delta_0$:
for each component, there is a panel corresponding to each value of $E$.
The curves correspond to different values of $\phi$ as indicated in the
legend. At $\phi=0$ and $\phi=180^\circ$ the same conservation
laws that preclude singlet to triplet pair conversion imply that the
torques vanish.  It is evident that there is no point in including
the regions of the sample deep inside $S$ or even 
well inside $F_1$, so the 
region plotted is that which includes both interfaces: the $S/F_2$ interface at the origin and that between ferromagnets at $Y=-30$, where $Y$ is the dimensionless position. 

The $y$-components results are easiest to understand: the component of the
torque has very sharp peaks, with opposite signs, near the $F_1/F_2$ boundary where it vanishes. These peaks reflect the existence
of a strong but short-ranged magnetic proximity effect. In $F_2$ and 
in $F_1$, $\tau_y$ is  small and oscillatory. It reaches its maximum value at $\phi=90^\circ$. It depends only weakly on the bias, since it basically
reflects a static effect: the two magnets interacting with each other. This
behavior is of course reflected in $S_y$ as both quantities are 
related via Eq.~(\ref{spinconserve}). 

The behavior of the in-plane components, $x$  and $z$, is similar to each other 
(they are related by spin rotations) and quite different  from that of $y$.
Now currents and torques are transport-induced and one sees immediately
that they markedly depend on bias. Since in $F_1$ the internal
field always points along $z$, we find that $S_z$ is a constant
in $F_1$, its value increasing with bias. As a function of $\phi$
its behavior is complicated, the maximum value is not precisely 
at  $\phi=90^\circ$ and it is dependent on bias. 
For this value of $\phi$ 
the field points along the $x$ direction in $F_2$ (it is always
along $z$ in $F_1$). Therefore $S_z$ is always spatially constant in $F_1$
and this applies also to $S_x$ in $F_2$ at $\phi=90^\circ$. 
For other values
of the mismatch angle $S_x$ oscillates in both magnetic layers, and so
does $S_z$ in $F_2$. The 
amplitude of the oscillations of $S_x$ decays 
 slowly deep into the $F_1$ layer. 
In all cases the period of the spatial
oscillations is approximately $1/h$ indicating that the oscillations are
due to the behavior of the Cooper pairs. As to the corresponding
components of the torque, one notes at once that their maximum value
is much smaller than that of the $\tau_y$ peak but, away from the $F_1/F_2$ interface, the values are not  all that different.
This reflects the geometry, as explained above. 
We see that the $x$ and $z$ components of the torque are also nonmonotonic with
$\phi$, with 
 peaks that are not necessarily at $\phi=90^\circ$, depending on the bias. For lower biases, the peak values appear to shift away 
to smaller values,  
more closely aligned with the $z$ direction, due to the increasing static effect from the $F_1$ layer. 
In our coordinate system, $\tau_z$
vanishes in $F_1$ for all $\phi$ and oscillates in $F_2$. 
Correspondingly, $\tau_x$ is oscillatory in both $F_1$ and $F_2$ except at $\phi=90^\circ$ where it is zero in $F_2$. 
We have not plotted the magnetization itself,
but its components exhibit damped oscillations which reflect the 
well known\cite{ralph}precessional 
behavior of the magnetization around the internal fields. Such precessional
behavior is then reflected in the current oscillations discussed above.

\begin{figure}
\hspace*{-20mm}\includegraphics[width=0.55\textwidth,angle=-90]{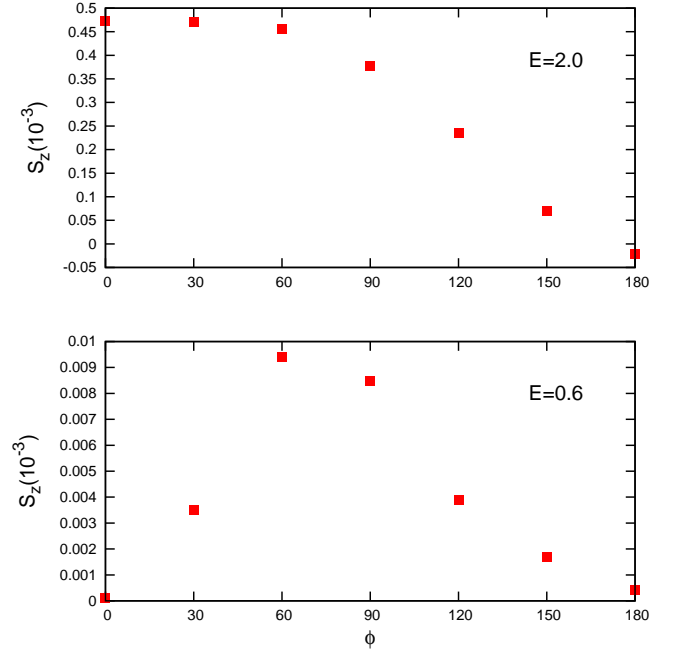}
\caption{The $z$ component of the spin current in the outer $F$ region
as a function of $\phi$, at two different bias values.}
\label{figure10}
\end{figure}

In our coordinate system, $S_z$ is a constant in the outer
layer, $F_1$. Also, all the components of the
spin current are trivially constant in the $S$ layer, since
there are no torques there. 
As can be seen in Fig.~\ref{figure8}, all spin
current components vanish in $S$ unless the bias exceeds
the bulk $S$ gap, $\Delta_0$. This confirms the remarkable
fact\cite{wvhg} that, in this respect, spin currents behave like 
charge currents in an $N/S$ junction. It can rather easily be shown via standard
spin rotation matrix arguments  that the constant values of $S_z$ and $S_x$
deep in the $S$ material, in the limit of  large bias, 
should be approximately related to the value of $S_z$
in  the $F_1$ layer  by factors of
$\cos \phi$ and $\sin \phi$ respectively, and this  can
be seen in the last column of Fig.~\ref{figure8} to hold rather
accurately at $E=2$. On the other hand, the dependence of the
constant value of $S_z$ in the outer layer on $\phi$  is nontrivial
as one can see in Fig.~\ref{figure8}. 
We display this more clearly in Fig.~\ref{figure10}, 
where we plot the value of $S_z$ in $F_1$ at two different bias values.
We see that for values below the CB the behavior is nonmonotonic: it
cannot be, since $S_z$ vanishes at both $\phi=0$ and $\phi=180^\circ$. The
maximum value is near $\phi=90^\circ$. On the other hand, when
the bias is well above the CB, $S_z$, which in this 
case is non-vanishing at zero angular mismatch, 
decreases monotonically with $\phi$. It becomes slightly
negative when the two magnets are aligned in opposite
direction. The
behavior is not described by a simple trigonometric function and 
a simple argument leading to the behavior 
found seems elusive. 

\section{Conclusions} 
\label{conclusions} 
The focus of this paper is on the prediction of 
the charge transport properties of superconducting spin valves 
with a $F_1/N/F_2/S$ layered structure. The emphasis is on
studying systems having material and geometrical
characteristics corresponding to samples that can realistically
be experimentally fabricated. Our main results 
pertain to the conductance $G$  as a function of bias, 
particularly with respect to the 
misalignment magnetization angle $\phi$ between the $F$ layers: variation of this angle produces the desired spin valve effects. 
The conductance is the basic information which is
experimentally obtained from charge transport measurements: it
is the derivative of the current-voltage relation.
To further our objective 
we have used values of the material parameters (such as the internal
magnetic field and  the superconducting coherence length)
which have been previously shown\cite{alejandro} to fit with
great accuracy the transition temperatures of such valve structures
when the actual materials are Co, Cu and Nb. We have also used thickness
values which encompass the available and desirable experimental 
ranges and have stayed away from idealistic assumptions, such as
ideal interfaces, which are essentially irrelevant to actual
experimental conditions. We have also studied the often neglected 
temperature dependence of the results. We have used a fully self
consistent approach, which is absolutely necessary to ensure
that charge conservation is satisfied.

Our results are summarized in Sect.~\ref{results}. The most important
conclusion to be learned from the figures 
presented is that simple
extrapolations are inadequate. There are several interfering oscillatory
phenomena involved -- the  center of mass oscillation of the Cooper pairs
in ferromagnets, the transmissions and reflections (ordinary,
Andreev, and anomalous Andreev) at the three interfaces, and the usual
quantum mechanical effects. As a result, the dependence of the 
relevant quantities that characterize the conductance (examples
are the critical bias, the zero bias conductance, and the low and high bias features) 
have nonmonotonic behavior when just about any parameter in the problem
varies. From this it follows that the valve effects, that is, the variation
of $G$ with $\phi$, vary quantitatively and 
qualitatively  depending 
on parameter values. 
The lack of monotonicity makes it extremely difficult to predict 
by extrapolation 
the measurable features expected 
for any given set of conditions. 
 The only thing that makes sense is to build
a database of conductance plots for different sets of
parameter values, and compare the plots in the database with experimental
results as they become available. We have
built such a database--
the results included here are a representative subset.

As far as the geometry dependence we have found that results depend
most strongly on the thickness of the inner ferromagnetic layer, 
with a large dependence on the
normal spacer thickness 
as well and a relatively weaker one on  
that of the outer $F$ electrode. 
This is however an overall, general statement: specific  details 
may be different. We have also found that the 
interfacial scattering specifically due to surface imperfections (the barriers),  does not severely affect the valve effects for typical experimentally accessible values. 
Of course, scattering
strong enough to destroy the proximity effect would be another matter.
Another important conclusion we have reached is that temperature effects
are not negligible in typical experimental
situations. 
Furthermore, 
because of high derivative regions in the $G$ vs. bias curves, a
Sommerfeld expansion does not work well. However, an exact calculation
can be performed numerically and it reveals that the shape of 
the conductance curve changes, becoming much smoother as bias varies, where as the valve effects as a function of $\phi$ remain unaffected. 

We have also studied, in a much more limited way, the spin transfer torque
and the spin currents in structures lacking the $N$ layer. The results are
analyzed in Sec.~\ref{spincurrents}. 
We have found, in our geometry, that the $y$-component of the spin torques have sharp peaks at the $F_1/F_2$ interface, nearly independent of applied bias. 
These are due to the strong, static magnetic proximity effects. 
The greatest peak occurs for a  mismatch angle $\phi$ of $90^\circ$. 
The spin torque components in the $x$ and $z$ direction are bias dependent %
and more complex, with higher peaks at angles smaller than $\phi=90^\circ$ 
for lower biases. We attribute this to static effects from the $F_1$ layer
 magnetization. We have calculated the spin currents using the direct method
 described in Eq.~\ref{spincur}. We find a nonmonotonic behavior in 
 the spin current amplitudes similar to that of the spin torque. 
 The oscillation amplitudes tend to peak for angles slightly below
  $\phi=90^\circ$ for lower biases. 
  The $S_z$ component is constant in the $F_1$ layer and  monotonic 
 with angle for high bias values (above $\Delta_0$) only. In the $S$ layer, 
 the spin currents are zero except for at high bias when both 
 the $S_x$ and $S_z$ components attain nonzero values for most values 
 of $\phi$. 
The consistency between
the torques and  spin current gradients, imposed by the conservation
laws, is ensured in our approach.

To conclude, the measurable quantities  have complex behavior,
often nonmonotonic as  experimental parameters and inputs
vary. Our plots provide  an wide spectrum of features to study, 
many of which are not yet fully understood. 
We expect that the results we have obtained will provide
a very important guide to experimentalists building real world superconducting
spin valves in nanoscale heterostructures.

\acknowledgments  The authors thank I.N. Krivorotov (University
of California, Irvine) for many
illuminating discussions on the
experimental issues. They are very grateful to Chien-Te Wu
(National Chiao Tung University) for many helpful
discussions on all aspects of this
problem. They also thank Yanjun Yang for technical 
help with the spin current
calculations. This work was supported in part by DOE grant No. DE-SC0014467

\end{document}